\documentclass[showpacs,twocolumn,prd,aps]{revtex4}
\usepackage{amsfonts}
\usepackage{amsmath}
\usepackage{amsbsy}
\usepackage{amssymb}

\begin{document}

\title{Equilibrium configurations of two charged
masses in General Relativity}
\author{G.A.~Alekseev$^1$}
  \email{G.A.Alekseev@mi.ras.ru}
\author{V.A.~Belinski$^2$}
  \email{belinski@icra.it}
\affiliation{\centerline{\hbox{$^1$Steklov Mathematical
Institute, Gubkina 8, Moscow 119991, Moscow, Russia}} \\
\centerline{\hbox{${}^2$
INFN, Rome University "La Sapienza",\,\,\, 00185 Rome, Italy,}}\\
\centerline{\hbox{${}$ ICRANet, Piazzale della Repubblica, 10, 65122 Pescara, Italy}}\\
\centerline{\hbox{and IHES, F-91440 Bures-sur-Yvette, France}}\\
}


\begin{abstract}
\noindent
An asymptotically flat static solution of Einstein-Maxwell equations which describes the field of two non-extreme  Reissner - Nordstr\"om sources in equilibrium is presented. It is expressed in terms of physical parameters of the sources (their masses, charges and separating distance). Very simple analytical forms were found for the solution as well as for the equilibrium condition which guarantees the absence of any struts on the symmetry axis. This condition shows that the equilibrium is not possible for two black holes or for two naked singularities. However, in the case when one of the sources is a black hole and another one is a naked singularity, the equilibrium is possible at some distance separating the sources. It is interesting that for appropriately chosen parameters even a Schwarzschild black hole together with a naked singularity can be "suspended"  freely in the superposition of their fields.
\end{abstract}


\maketitle

\subsection*{Introduction}
   In the Newtonian physics two point-like particles can be in equilibrium if the product of their masses is equal to the product of their charges (we use the units for which $G=c=1$). In General Relativity, till now the equilibrium condition for two particle-like sources imposed on their physical masses, charges and separating distance was not known in explicit and reasonably simple analytical form which would  admit a rigorous analysis without a need of numerical experiments. The only exceptional case was the Majumdar-Papapetrou solution \cite{Majumdar:1947, Papapetrou:1947}, for which the charge of each source is equal to its mass. In this case, the equilibrium is independent of the distance between the sources. For each of the static sources of this sort  its outer and inner Reissner-Nordstr\"om horizons coincide and such sources are called extreme ones. Accordingly, the sources with two separated horizons are called under-extreme and the sources without horizons -- super-extreme.

   The problem, which had been under investigation by many researchers and  which we solve in the present paper, consists in the search of  equilibrium configurations of non-extreme sources. Since the advent of  solution generating techniques for stationary axisymmetric Einstein-Maxwell fields, a construction of an exact solution for two charged masses at rest does not represent any principal difficulty. However, in general the asymptotically  flat  solutions of this kind contain conical singularities on the symmetry axis between the sources which can be interpreted as a presence of some extraneous struts preventing the sources to fall onto or to run away from each other. The equilibrium condition just implies the absence of such struts. Naturally, if the metric is known so is the equilibrium condition. In the static case,  the latter means that the product of the metric coefficients $g_{tt}$
and $g_{\rho\rho}$  (in cylindrical Weyl coordinates) should be equal to unity at the axis  where  $\rho=0$. However, this equilibrium equation in such general form usually is expressed by a set of formal parameters and it is so complicated that its analytical investigation appears to be very difficult. Therefore, it is desirable to have this equation expressed in terms of physical parameters and in a simple enough form making it accessible for an analytical examination of a possibility of realization of equilibrium. Moreover, this realization should be compatible with a condition of a positive value of the distance between the sources. This task have not  been accomplished yet, and up to now there were known  only some results achieved by numerical calculations.

   The first researches of the equilibrium of non-extreme sources \cite{Barker-O'Connell:1977} - \cite{Azuma-Koikawa:1994} led to the contradictory conclusions. The authors of the indicated papers used both the exact techniques and pN and ppN approximations. The common opinion
expressed in \cite{Barker-O'Connell:1977, Kimura-Ohta:1977} and \cite{Ohta-Kimura:1982} - \cite{Azuma-Koikawa:1994} was that the equilibrium for non-extreme sources is impossible. Nevertheless, in \cite{Tomimatzu:1984} one can find a remark that the analysis performed was insufficient and the existence of equilibrium configurations for the non-extreme objects can not be excluded. The arguments in favour of such possibility can be found also in \cite{Bonnor:1981}.

   The next step which attracted attention to the problem again have been done by Bonnor in \cite{Bonnor:1993}, where the equilibrium condition for a charged test particle in the Reissner-Nordstr\"om field was analyzed. Examination made there suggested also some plausible assumptions for the exact solutions. As have been indicated in \cite{Bonnor:1993} a charged test body can be at rest  in the field of the Reissner-Nordstr\"om source only if they both are either extreme (for the test particle the degree of its extremality is defined just by the ratio between its charge and mass), balanced irrespective of distance, or one of them is super-extreme and the other is under-extreme, and in this case the equilibrium depends on the distance. There is no way for equilibrium in cases when both sources are either super-extreme or under-extreme. It is worth to mention that in the very recent papers \cite{Bini-Geralico-Ruffini:2007} a new perturbative solution describing an equilibrium state of two-body system consisting of a Reissner-Nordstr\"om black hole and a super-extreme test particle has been presented. The whole set of combined Einstein-Maxwell equations has been solved there by using the first order perturbation approach developed in \cite{Zerilli:1974} and based on the tensor harmonic expansion of both the gravitational and electromagnetic fields adopting the Regge-Wheeler \cite{Regge-Wheeler:1957} gauge. (The basic equations for combined  gravitational and electromagnetic perturbations of the Reissner-Nordstr\"om background in the decoupled form  were found in another gauges in \cite{Sibgatullin-Alekseev:1974} and in the also decoupled   Hamiltonian form in \cite{Moncrief:1974}).  Both the electromagnetically induced gravitational perturbations and  gravitationally induced electromagnetic perturbations  \cite{Johnston-Ruffini-Zerilli:1973} due to a mass as well as a charge of the particle have thus taken into account.  The expressions in a closed form for both the perturbed metric and electromagnetic field have been explicitly given \cite{Bini-Geralico-Ruffini:2007}. It is interesting that the equilibrium equation (which arises in this case as a self-consistency condition for the set of differential equations for perturbations) remains the same as of Bonnor \cite{Bonnor:1993}.

   The Bonnor's  analysis allows to expect that qualitatively the same can happen also for two  Reissner-Nordstr\"om sources. For two extreme sources this is indeed the case because it is known that such generalization exists and leads to the Majumdar-Papapetrou solution. Up to 1997 it remained unknown whether the analogous generalization for the non-extreme bodies can be found. The first solid arguments in favour of existence of a static equilibrium configuration for the  "black hole - naked singularity" system was presented in \cite{Perry-Cooperstock:1997}. These results have been obtained thereby numerical calculations and  three examples of  numerical solutions of the equilibrium equation have been demonstrated. These solutions can correspond to the equilibrium configurations free of struts. For the complete proof it would be necessary to show that such configurations indeed consist of two sources, separated by physically sensible distance between them. However, in \cite{Perry-Cooperstock:1997} it was pointed out that the distance dependence for the equilibrium state is unknown. The authors of \cite{Perry-Cooperstock:1997} also reported that a number of numerical experiments for two black holes and for two naked singularities showed the negative outcomes, i.e. all tested sets of parameters were not in power to satisfy the equilibrium equation. These findings are in agreement with Bonnor's  test particle analysis. One year later the similar numerical analysis was made in \cite{Breton-Manko-Sanches:1998}.

   In this paper, we present an exact solution of the Einstein-Maxwell equations which describes the field of two Reissner-Nordstr\"om sources in static equilibrium as well as the equilibrium condition itself which turns out to have unexpectedly simple form expressed in terms of physical parameters of the sources. This simplicity permits us to prove a validity of conjectures of the papers \cite{Bonnor:1993} and \cite{Perry-Cooperstock:1997}  on exact analytical  level. It allows also a direct analytical investigation of the physical properties of the equilibrium state of two non-extreme sources.

   We precede a description of our results with a few words on the methodology of  the derivation of our solution. An application for derivation of this solution of the Inverse Scattering Method for electro-vacuum  developed in \cite{Alekseev:1980, Alekseev:1988} and described  in details in the book \cite{Belinski-Verdaguer:2001} leads to not most convenient parametrization of the solution which give rise to some subsequent technical difficulties (although there are no principal obstacles to use this approach). Instead, we used the Integral Equation Method  \cite{Alekseev:1985, Alekseev:1988} which opens a shorter way to the desirable results. The first step was to construct the solution for
the two-pole structure of the monodromy data on the spectral plane with a special choice of parameters providing asymptotical flatness and the static character of the solution. This corresponds also to the two-pole structure of the Ernst potentials (as functions of the Weyl cylindrical coordinate z) on the symmetry axis. Then the expressions for physical masses and physical charges for both sources were found with the help of the Gauss theorem and the notion of distance between these sources was also defined. We stress here that the physical character of masses and charges of the sources follows not only from their definition using the Gauss theorem, but also from the analysis of that limiting case in which one of the sources is a test particle (see the formulae (12), (13) below and the text after them). After that we derived the equilibrium equation in terms of these five physical parameters. The miracle arises if one substitutes this equilibrium equation back into the solution. This  results in the impressive simplification of all formulas. Below we expose the final outcome which is ready for using in practical purposes without necessity of knowledge of any details of its derivation.

It is worthwhile to mention that a correctness of our solution has been confirmed also by its direct substitution into the Einstein - Maxwell field equations.

\section*{The solution}

For our static solution in cylindrical Weyl coordinates,
metric and vector electromagnetic potential take the forms
\begin{eqnarray}
ds^2 &=& H dt^2-f(d\rho^2+dz^2)-\dfrac{\rho^2}{H} d\varphi^2,\label{Metric} \\
A_t &=& \Phi,\qquad A_\rho=A_z= A_\varphi=0,
\label{Potential}\end{eqnarray}
where $H$, $f$ and $\Phi$ are real functions of the coordinates $\rho$ and $z$. These functions take the most simple form in bipolar coordinates which consist of two pairs of spheroidal variables $(r_1,\theta_1)$, $(r_2,\theta_2)$ defined by their relations to the Weyl coordinates:
\begin{equation}
\begin{array}{l}
\left\{\begin{array}{lccl}
\rho=\sqrt{(r_1-m_1)^2-\sigma_1^2}\sin\theta_1,\\[1ex]
z=z_1+(r_1-m_1)\cos\theta_1,
\end{array}\right.\\[4ex]
\left\{\begin{array}{lccl}
\rho= \sqrt{(r_2-m_2)^2-\sigma_2^2}\sin\theta_2,\\[1ex]
z=z_2+(r_2-m_2)\cos\theta_2.
\end{array}\right.
\end{array}
\end{equation}
Here and below, the indices ${}_1$ and ${}_2$ denote the coordinates and parameters related to the Reissner - Nordstr\"om sources located at the symmetry axis at the points $z=z_1$ and  $z=z_2$ respectively. A positive constant $\ell$ defined as
\begin{equation}\label{distance}
\ell=z_2-z_1
\end{equation}
characterizes the z-distance separating these sources (for definiteness we take $z_2 > z_1$). The constants $m_1$ and $m_2$ are physical masses of the sources.

   Each of the parameters $\sigma_k$ ($k=1,2$)                          can be either real or pure imaginary and this  property characterizes the corresponding Reissner - Nordstr\"om source to be either a black hole or a naked singularity: the real value of $\sigma_k$ means that this is a black hole whose horizon in Weyl coordinates is $\{\rho=0,\,z_k-\sigma_k\le z\le z_k+\sigma_k\}$ while the imaginary $\sigma_k$  corresponds to a naked singularity whose critical spheroid  $r_k=m_k$ is $\{0\le \rho\le\vert\sigma_k\vert,\, z=z_k\}$.
So the coordinate distance between two black holes (both $\sigma_1$ and $\sigma_2$ are real and positive) we define as the distance along z-axis between  the nearest points of its intersections with two horizons and this distance is $\ell-\sigma_1-\sigma_2$. The distance between the black hole located at the point $z=z_2$  and the naked singularity  at the point $z=z_1$ ( $\sigma_2$ is real and positive but $\sigma_1$ is pure imaginary) we define as distance between the nearest points of intersections of the symmetry axis with black hole horizon and critical spheroid and this distance is $\ell-\sigma_2$. The distance between two naked singularities (both $\sigma_1$ and $\sigma_2$ are pure imaginary) is simply  $\ell$  and it is the length of the segment between the nearest points of intersections of two critical spheroids with the z-axis.

 In terms of  bipolar coordinates our solution reads:
\begin{eqnarray}
\label{gtt} H &=&[(r_1-m_1)^2-\sigma_1^2+\gamma^2\sin^2\theta_2]\\
\nonumber &&\times[(r_2-m_2)^2-\sigma_2^2+\gamma^2\sin^2\theta_1]
\mathcal{D}^{-2},\\
\label{At}\Phi &=&[(e_1-\gamma)(r_2-m_2)+(e_2+\gamma)(r_1-m_1)\\
\nonumber &&+\gamma(m_1\cos\theta_1+m_2\cos\theta_2)]
\mathcal{D}^{-1},  \\
\label{Factorf}f&=&[(r_1-m_1)^2-\sigma_1^2\cos^2\theta_1]^{-1}\\
\nonumber&&\times
[(r_2-m_2)^2-\sigma_2^2\cos^2\theta_2]^{-1}
\mathcal{D}^2,
\end{eqnarray}
where
\begin{equation}\label{Determinant}
    \mathcal{D}=r_1 r_2-(e_1-\gamma-\gamma \cos\theta_2)
(e_2+\gamma-\gamma \cos\theta_1).
\end{equation}
In these expressions the quantities $e_1$, $e_2$ represent physical charges of the sources. The parameter $\gamma$  and the parameters
$\sigma_1$, $\sigma_2$ are determined by the relations:
\begin{equation}\label{Sigmas}
\begin{array}{l}
  \sigma_1^2=m_1^2-e_1^2+2 e_1\gamma, \quad
  \sigma_2^2=m_2^2-e_2^2-2 e_2\gamma,\\[1ex]
\gamma=(m_2 e_1-m_1 e_2)(\ell+m_1+m_2)^{-1}.
\end{array}
\end{equation}
The formulas (\ref{Metric})-(\ref{Sigmas}) give the exact solution of the Einstein-Maxwell equations if and only if the five parameters $m_1$, $m_2$, $e_1$, $e_2$ and $\ell$ satisfy the following condition
\begin{equation}\label{Equilibrium}
    m_1 m_2=(e_1-\gamma)(e_2+\gamma).
\end{equation}
The condition (\ref{Equilibrium}) guarantees the equilibrium without any struts on the symmetry axis between the sources.

\section*{Properties of the solution}
First of all one can see that the balance equation (\ref{Equilibrium}) do not admit two black holes ($\sigma_1^2>0$, $\sigma_2^2>0$) to be in equilibrium under the condition that there is some distance between them, that is if $\ell-\sigma_1-\sigma_2>0$. This is in an agreement with a non-existence of static equilibrium configurations of charged black holes proved under rather general assumptions in \cite{Chrusciel-Tod:2007}. (To avoid a confusion, we mention here that the results of \cite{Chrusciel-Tod:2007} do not apply in the presence of naked singularities.) The equilibrium is also impossible if one of the sources is extreme and the other is a non-extreme one and a positive distance exists between them, i.e. if $\ell-\sigma_2>0$ for the case $\sigma_1=0$ and $\sigma_2^2>0$ (a negative value for $\sigma_2^2$       is forbidden at all if $\sigma_1=0$)\,\,\footnote{Non-separated objects for which the horizons overlap each other or the horizon intersects with the critical spheroid also may be possible but such cases are not in the scope of this communication.}.
The condition (\ref{Equilibrium}) implies also that $\sigma_1^2$ and $\sigma_2^2$ never can be both negative, that is the equilibrium of two naked singularities is impossible. So, for separated sources an equilibrium may exist  either between a black hole and a naked singularity or between two extreme sources. The latter case can be realized only if $\sigma_1=\sigma_2=0$, $\gamma=0$ and it is easy to see that the formulas (\ref{Metric})-(\ref{Sigmas}) reduce for this case to the Majumdar-Papapetrou solution.

At spatial infinity the variables $r_1$, $r_2$ coincide and one can choose any of them as the radial coordinate. In this region the fields, as can be seen from (\ref{gtt}) and (\ref{At}), acquire the standard Reissner-Nordstr\"om asymptotical form with the total mass $m_1+m_2$ and the total charge $e_1+e_2$.

At the symmetry axis $\cos^2\theta_1=\cos^2\theta_2=1$ and the formulas (\ref{gtt}), (\ref{Factorf}) show that the condition $f H=1$ is satisfied there automatically, i.e. there are no conical singularities. Besides the singularities inherent to the sources themselves, any other kind of singularities (such as, for example, the off-axis singularities found in the double-Kerr solution in \cite{Bicak-Hoenselaers:1985}) are also absent in our solution.

The constant $\gamma$ vanishes in the limit $\ell\to\infty$  whence it follows from (\ref{Equilibrium}) that the equilibrium condition asymptotically reduces to the Newtonian form $m_1 m_2=e_1 e_2$ for a large distance between the sources.

If one of the sources disappears, e.g. $m_1=e_1=0$, our solution reduces to the exact Reissner-Nordstr\"om solution with the mass $m_2$ and the charge $e_2$  in the standard spherical coordinates $r_2$, $\theta_2$.

Let us turn now to that limiting case in which one of the sources can be considered as a test particle. For this we assume that $m_1$ and $e_1$ are infinitesimally small but the ratio $e_1/m_1$ is finite. In this case, in the first non-vanishing order with respect to the constants $m_1$ and $e_1$ the equilibrium condition (\ref{Equilibrium}) gives:
\begin{equation}\label{Test}
    (\ell+m_2)(m_1 m_2-e_1 e_2)=(m_1 e_2-m_2 e_1) e_2.
\end{equation}
We introduce instead of $m_1$  a new parameter $\mu_1$ defined by the relation:
\begin{equation}\label{Defmu}
\begin{array}{l}
    m_1=\mu_1[1-2 m_2(\ell+m_2)^{-1}+e_2^2(\ell+m_2)^{-2}]^{1/2}\\[1ex]
    \phantom{m_1=}+e_1 e_2(\ell+m_2)^{-1}.
\end{array}
\end{equation}
Now the relation (\ref{Test}) takes the form:
\begin{equation}\label{Testequilibrium}
\begin{array}{l}
m_2-e_2^2(\ell+m_2)^{-1}\\
 =e_1 e_2\mu_1^{-1}[1-2 m_2(\ell+m_2)^{-1}+e_2^2(\ell+m_2)^{-2}]^{1/2}.
\end{array}
\end{equation}
This last equation is nothing else but the Bonnor's balance condition \cite{Bonnor:1993} for the test particle of the rest mass $\mu_1$ and the charge $e_1$ in the Reissner-Nordstr\"om field of the mass $m_2$ and the charge $e_2$. The particle is at rest on the symmetry axis at the point $R=\ell+m_2$ where $R$ is the radius of the standard spherical coordinates of the Reissner-Nordstr\"om solution.
If we calculate from (\ref{At})  the potential $\Phi$ in the linear approximation with respect to the small parameters $m_1$ and  $e_1$  for the particular case $e_2=0$ (i.e. for the Schwarzschild background) the result will coincide exactly with the potential which have been found first in  \cite{Hanni-Ruffini:1971} -  \cite{Hanni-Ruffini:1973} in the form of multipole expansion and then in \cite{Linet:1976} in closed analytical form.

The relation (\ref{Defmu}) is important  since it exhibits clearly the physical nature of the mass $m_1$ and gives its correct interpretation. This relation shows that the parameters $m_1$, $m_2$ are not the rest masses but they represent the total relativistic energy of  each source in the external field produced by its partner.

Finally it is worth to mention that our exact solution remains physically sensible also in the case $e_2=0$. This corresponds to a Schwarzschild black hole of the mass $m_2$ hovering freely in the field of a naked singularity of the mass $m_1$ and the charge $e_1$.  Such configuration exists due to the repulsive nature of gravity in the vicinity of the naked Reissner-Nordstr\"om singularity.

                                                                                              \subsection*{Acknowledgements}

   GAA is thankful to ICRAnet for the financial support and hospitality during his visit to ICRAnet (Pescara, Italy) during May 2006, when this paper was started. The work of GAA was also supported in parts by the Russian Foundation for Basic Research (grants 05-01-00219, 05-01-00498, 06-01-92057-CE) and the programs "Mathematical Methods of Nonlinear Dynamics" of the Russian Academy of Sciences, and "Leading Scientific Schools" of Russian Federation (grant NSh-4710.2006.1).

We are especially grateful to R. Price for useful comments which  urged us to improve essentially this paper.


\begin{thebibliography}{99}

\bibitem{Majumdar:1947} S. D. Majumdar, Phys.Rev., {\bf 72}, 390 (1947).
\bibitem{Papapetrou:1947} A. Papapetrou, Proc.Roy.Irish Academy, Dublin, {\bf LI}, 191 (1947).
\bibitem{Barker-O'Connell:1977} B. M. Barker and R. F. O'Connell, Phys. Lett.,  {\bf 61A}, 297 (1977).
\bibitem{Kimura-Ohta:1977} T. Kimura and T. Ohta, Phys. Lett.,
{\bf 63A}, 193 (1977).
\bibitem{Bonnor:1981} W. B. Bonnor, Phys. Lett., {\bf 83A}, 414 (1981).
\bibitem{Ohta-Kimura:1982} T. Ohta and T. Kimura, Prog. Theor. Phys.,
{\bf 68}, 1175 (1982).
\bibitem{Tomimatzu:1984} A. Tomimatzu, Prog. Theor. Phys., {\bf 71}, 409 (1984).
\bibitem{Azuma-Koikawa:1994} T. Azuma and T. Koikawa, Prog. Theor. Phys., {\bf 92}, 1095 (1994).
\bibitem{Bonnor:1993} W. B. Bonnor, Class. Quant. Grav., {\bf 10}, 2077 (1993).
\bibitem{Bini-Geralico-Ruffini:2007} D. Bini, A. Geralico, R.J. Ruffini, Phys. Rev. {\bf D75}, 044012 (2007); Phys. Lett.,  {\bf 360A}, 515 (2007).
\bibitem{Zerilli:1974} F. J. Zerilli, Phys. Rev. {\bf D9}, 860 (1974).
\bibitem{Regge-Wheeler:1957}
T. Regge and J. A. Wheeler, Phys. Rev. {\bf 108}, 1063 (1957).
\bibitem{Sibgatullin-Alekseev:1974} N.R.Sibgatullin and G.A.Alekseev, Zh. Eksp. Teor. Fiz., {\bf 67}, 1233 (1974); [Sov. Phys. JETP, {\bf 40}, 613 (1975)].
\bibitem{Moncrief:1974} V. Moncrief, Phys. Rev. {\bf D9}, 2707 (1974); Phys. Rev. {\bf D10}, 1057 (1974).
\bibitem{Johnston-Ruffini-Zerilli:1973} M. Johnston, R. Ruffini, and F. J. Zerilli, Phys. Rev. Lett. {\bf 31}, 1317 (1973);
Phys. Lett. {\bf B49}, 185 (1974).
\bibitem{Perry-Cooperstock:1997} G. P. Perry and F. I. Cooperstock, Class. Quant. Grav., {\bf 14}, 1329 (1997).
\bibitem{Breton-Manko-Sanches:1998} N. Breton, V. S. Manko and J. A. Sanches, Class. Quant. Grav., {\bf 15}, 3071 (1998).
\bibitem{Alekseev:1980} G. A. Alekseev, JETP Lett., {\bf 32}, 277 (1980).
\bibitem{Alekseev:1988} G.A.Alekseev, Proc. Steklov Math. Inst., Providence, RI: American Math. Soc., {\bf 3}, 215 (1988).
\bibitem{Belinski-Verdaguer:2001} V. Belinski and E. Verdaguer, Gravitational Solitons, CUP, Cambridge  (2001).
\bibitem{Alekseev:1985} G.A.Alekseev, Sov.Phys.Dokl., {\bf 30}, 565 (1985).
\bibitem{Chrusciel-Tod:2007} P. Chrusciel and P.Tod, Commun.Math.Phys., {\bf 271} 577 (2007); gr-qc/0512043
\bibitem{Bicak-Hoenselaers:1985} J.Bi$\check{\text{c}}$$\acute{\text{a}}$k and C.Hoenselaers, Phys. Rev. {\bf D31}, 2476 (1985).
\bibitem{Hanni-Ruffini:1971} R.S.Hanni and R.Ruffini, Bull.Am.Phys.Soc., {\bf 16}, 34, (1971).
\bibitem{Cohen-Wald:1971} J.Cohen and R.Wald, J.Math.Phys., {\bf 12}, 1845 (1971).
\bibitem{Hanni-Ruffini:1973} R. Hanni and R. Ruffini,
Phys. Rev. {\bf D8}, 3259 (1973).
\bibitem{Linet:1976} B. J. Linet, J. Phys. A: Math. Gen. {\bf 9}, 1081 (1976).
\end{thebibliography}
\end{document}